\documentclass[namedreferences]{SolarPhysics}
\usepackage[optionalrh,solaenum]{spr-sola-addons} % For Solar Physics
\usepackage{subfig}
\usepackage{caption}
\usepackage{graphicx}        % For eps figures, newer & more powerfull
 \usepackage{color}           % For color text: \color command
\usepackage{url}             % For breaking URLs easily trough lines
 \usepackage{amssymb}
\newcommand{\Rmnum}[1]{\expandafter\@slowromancap\romannumeral #1@}

\begin{document}
\begin{article}
\begin{opening}
%\title{Full-Disk Nonlinear Force-Free Field Reconstruction of SDO/HMI Magnetogram in Spherical Geometry}
\title{A Comparison Between Nonlinear Force-Free Field and Potential Field Models Using Full-Disk SDO/HMI Magnetogram}
\author{Tilaye~\surname{Tadesse}$^{1}$,%\,$^{2}$,
         T.~\surname{Wiegelmann}$^{3}$, 
           B.~\surname{Inhester}$^{3}$,       
              K.~\surname{Olson}$^{1}$,
       P. J. ~\surname{MacNeice}$^{4}$, 
            A.~\surname{Pevtsov}$^{5}$
        }
\runningauthor{T.~Tadesse et al.}
%\runningtitle{Full-Disk Nonlinear Force-Free Field Reconstruction of SDO/HMI Magnetogram}
\runningtitle{A Comparison Between Nonlinear Force-Free Field and Potential Field Models}
 \institute{$^{1}$ Department of Physics, Drexel University, Philadelphia, PA 19104-2875, U.S.A,
                        email: \url{tasfaw@einstein.physics.drexel.edu},email: \url{Kevin.M.Olson@drexel.edu} \\
%                  $^{2}$Addis Ababa University, College of Natural Sciences, Institute of Geophysics, Space Science, and Astronomy,
%                        Po.Box 1176, Addis Ababa, Ethiopia, 
%                         email: \url{tilaye.tadesse@gmail.com} \\
                $^{3}$ Max Planck Institut f\"{u}r Sonnensystemforschung, Max-Planck Str. 2, D--37191 Katlenburg-Lindau, Germany,
                      email: \url{wiegelmann@mps.mpg.de}, email: \url{inhester@mps.mpg.de}\\
                 $^{4}$ NASA, Goddard Space Flight Center, Code 674, Greenbelt, MD 20771, U.S.A.
                     email: \url{peter.j.macneice@nasa.gov } \\
                $^{5}$ National Solar Observatory, Sunspot, NM 88349, U.S.A.
                     email: \url{apevtsov@nso.edu} \\
    }      
            % }

\begin{abstract}
Measurements of magnetic fields and electric currents in the pre-eruptive corona are crucial to study 
solar eruptive phenomena, like flare and coronal mass ejections(CMEs). However, spectro-polarimetric 
measurements of certain photospheric lines permit a determination of the vector magnetic field at the photosphere. 
Thus, substantial collection of magnetograms relate to the photospheric surface field only. Therefore, there is considerable 
interest in accurate modeling of the solar coronal magnetic field using photospheric vector magnetograms 
as boundary data. This numerical modeling is carried out by applying state-of-the-art nonlinear 
force-free field (NLFFF) reconstruction. Cartesian nonlinear force-free field (NLFFF) codes are not 
well suited for larger domains, since the spherical nature of the solar surface cannot be neglected 
when the field of view is large. One of the most significant results of Solar Dynamic Observatory (SDO) 
mission to date has been repeated observations of large, almost global scale events in which large scale 
connection between active regions may play fundamental role. Therefore, it appears prudent to implement a 
NLFFF procedure in spherical geometry for use when large scale boundary data are available, such as from the Helioseismic 
and Magnetic Imager (HMI) on board SDO. In this work, we model the coronal magnetic field above multiple active 
regions with the help of a potential field and a NLFFF extrapolation codes in a full-disk using HMI data as 
a boundary conditions. We compare projections of the resulting magnetic field lines solutions with full-disk 
coronal images from the Atmospheric Imaging Assembly (SDO/AIA) for both models. This study has found that the NLFFF 
model reconstructs the magnetic configuration better than the potential field model. We have concluded that much of 
trans-equatorial loops connecting the two solar hemispheres are current-free. 

\end{abstract}
\end{opening}

%==============================================================================================================================================
\section{Introduction}
%==============================================================================================================================================
Magnetic fields are believed to play a dominant role for active phenomena carried out in the solar corona. The study of 
solar eruptive phenomena requires that we understand how magnetic energy is stored in the pre-eruptive corona. One of 
the key questions is: What is the three dimensional (3D) structure of magnetic fields and electric currents in the 
pre-eruptive corona, and how much free energy is stored in the field? \cite{schrijver:etal2005,Jiang:2012}. To answer 
this question we must measure the magnetic field in the coronal volume. However, spectro-polarimetric measurements of certain
photospheric lines permit a determination of the vector magnetic field at the  photosphere. Thus, substantial collection 
of magnetograms relate to the photospheric surface field only. Although measurement of magnetic fields in the chromosphere 
and the corona has considerably improved in recent decades \cite{Lin:2000,Liu:2008}, further developments are needed 
before accurate data are routinely available. The problem of measuring the coronal field and its embedded 
electrical currents thus leads us to use numerical modeling to infer the field strength in the higher layers of the solar 
atmosphere from the measured photospheric field. Due to the low value of the plasma $\beta$ (the ratio of gas pressure to 
magnetic pressure), the solar corona is magnetically dominated \cite{Gary}. To describe the equilibrium structure of the 
static coronal magnetic field when non-magnetic forces are negligible, the force-free assumption is appropriate:
\begin{equation}
   (\nabla \times\textbf{B})\times\textbf{B}=0 \label{one}
\end{equation}
\begin{equation}
    \nabla \cdot\textbf{B}=0 \label{two}
 \end{equation}
 subject to the boundary condition
\begin{equation}
    \textbf{B}=\textbf{B}_{\textrm{obs}} \quad \mbox{on photosphere} \label{three}
 \end{equation}
where $\textbf{B}$ is the magnetic field and $\textbf{B}_{\textrm{obs}}$ is measured vector field on the photosphere. 
Equation~(\ref{one}) states that the Lorentz force vanishes (as a consequence of $\textbf{J}\parallel \textbf{B}$, where 
$\textbf{J}$ is the electric current density) and Equation~(\ref{two}) describes the absence of magnetic monopoles. 

As an alternative to real measurements, force-free extrapolation of photosphericmagnetic fields is currently being used as a 
the primary tool for the modeling of coronal magnetic
fields \cite{Inhester06,valori05,Wiegelmann04,Wheatland04,Wheatland:2009,tilaye09,Wheatland:2011,Amari:2010,Wiegelmann:2012,Jiang:2012,Jiang:2012A,Aschwanden:2012,Wiegelmann:2012W}. 

Cartesian nonlinear force-free field (NLFFF) codes are not well suited for larger domains, since the spherical nature of 
the solar surface cannot be neglected when the field of view is 
large \cite{Wiegelmann07,tilaye09,Tilaye:2012a,Guo:2012}. One of the most significant results of Solar Dynamic Observatory (SDO) mission 
to date has been repeated observations of large, almost global scale events in which large scale connection between active regions 
may play fundamental role. Therefore, it appears prudent to implement a NLFFF procedure in spherical geometry for use when large scale 
boundary data are available, such as from the Helioseismic and Magnetic Imager (HMI) on board SDO. \inlinecite{DeRosa} has studied that 
different NLFFF models have markedly different field line configurations and provided widely varying estimates of the magnetic free 
energy in the coronal volume. The main reasons for that problem are (1) the forces acting on the field within the photosphere, (2) the 
uncertainties on vector-field measurements, particularly on the transverse component, and (3) the large domain that needs to be modeled 
to capture the connections of an active region to its surroundings \cite{Wiegelmann07,Tilaye:2010,Tilaye:2012,Tadesse:2013}. In this study, we have considered 
those three points explicitly into account for modeling coronal magnetic field in a full-disk.

In this work, we apply our spherical NLFFF procedure to a group of active regions observed on November 09 2011 around 17:45UT by 
SDO/HMI. During this observation, there were four active regions (ARs 11338, 11339, 11341 and 11342) along with other smaller 
sunspots spreading on the disk. We compare the extrapolated (both potential and NLFFF) magnetic loops with extreme ultraviolet (EUV) 
observations by the Atmospheric Imaging Assembly (AIA) \cite{Lemen:2012} on board SDO. During comparison, we check whether the NLFFF 
model reconstructs the magnetic configuration better than the potential field model. This comparison can be used to evaluate how well 
the model field lines approximate the observed coronal loops. 

%==============================================================================================================================================
\section{Instrumentation and data set}
%==============================================================================================================================================
The Helioseismic and Magnetic Imager (HMI) \cite{Schou:2012} is part of the Solar Dynamics Observatory (SDO)
and observes the full Sun at six wavelengths and full Stokes profile in the Fe {\Rmnum{1}} 617.3 nm spectral line. It 
consists of a refracting telescope, a polarization selector, an image stabilization system, a narrow band tunable filter 
and two 4096 pixel CCD cameras with mechanical shutters and control electronics. 

The transverse components of vector magnetic fields suffer from the so-called $180^{\circ}$ ambiguity. The $180^{\circ}$ ambiguity for
the HMI data in this study has been resolved by an improved version of the minimum energy method \cite{Metcalf:1994,Metcalf:2006,Leka:2009}. 
As described in \inlinecite{Leka:2009}, in weak-field areas, the minimization may not return a good solution due to large noise. 
The noise level is $\approx$ 10G and $\approx$ 100G for the longitudinal and transverse magnetic field, respectively.
Therefore, in order to get a spatially smooth solution in weak-field areas, we divide the magnetic field into two regions, i.e., 
strong-field region and weak-field region, which is defined to be where the field strength is below 200 G at the disk center, and 
400 G on the limb. The values vary linearly with distance from the center to the limb. The ambiguity solution in the strong-field 
area is derived by the annealing in the minimization method and released in the data series ``hmi.B\_720s\_e15w1332\_cutout``. 
Magnetic field azimuths in the weak-field area are finally determined by the potential-field acute-angle method after the annealing. This is
different from the original minimum energy method, which uses a neighboring-pixel acute-angle algorithm to revisit the weak field
area. Figure~\ref{fig3}a) shows the vector magnetic fields observed by HMI on November 09 2011 around 17:45UT. 
%==============================================================================================================================================
\section{Optimization principle and preprocessing of HMI data}
%==============================================================================================================================================
\inlinecite{Molodenskii69} and \inlinecite{Aly89} pointed out that vector magnetic fields on a closed surface that fully encloses 
any force-free domain have to satisfy the force-free and torque-free conditions. To serve as suitable lower boundary condition 
for a force-free modeling, on average a net tangential force acting on the boundary and shear stresses along axes lying on the 
boundary have to reduce to zero \cite{Molodenskii69,Aly89,Wiegelmann06sak,tilaye09}. Preprocessing method as implemented in 
\inlinecite{Wiegelmann06sak} has to be used to fulfill those conditions. The preprocessing scheme of \inlinecite{tilaye09} 
involves minimizing a two-dimensional functional of quadratic form in spherical geometry similar to
 \begin{displaymath} \textbf{B}=\emph{argmin}(L_{p}),
\end{displaymath}
\begin{equation}
L_{p}=\mu_{1}L_{1}+\mu_{2}L_{2}+\mu_{3}L_{3}+\mu_{4}L_{4},\label{3}
\end{equation}
where $\textbf{B}$ is the preprocessed surface magnetic field from the input observed field $\textbf{B}_{obs}$. Each of the constraints 
$L_{n}$ is weighted by an as yet undetermined factor $\mu_{n}$. The first term $(n=1)$ corresponds to the force-balance condition, the 
next $(n=2)$ to the torque-free condition, and the last term $(n=4)$ controls the smoothing. The explicit form of $L_{1}$, $L_{2}$, and 
$L_{4}$ can be found in \inlinecite{tilaye09}. The term $(n=3)$ controls the difference between measured and preprocessed vector fields. 
After preprocessing the HMI data, we solve the force-free equations (\ref{one}) and (\ref{two}) using optimization principle \cite{Wheatland00,Wiegelmann04} 
in spherical geometry \cite{Wiegelmann07,tilaye09,Tadesse:2013}. 
%==============================================================================================================================================
\section{Potential field model}
%==============================================================================================================================================
The simplest way to model the coronal field is to assume that it is potential, i.e. that it carries no electric current. 
Solutions for this model in plane geometry have been obtained by \inlinecite{schmidt}, for the case where the vertical 
component of the field is specified at the photospheric boundary. Potential field model has now led to an almost routine type 
reconstruction, used for observational purposes \cite{Sakurai:1989}, but also for building initial conditions
for dynamical MHD numerical simulations \cite{Amari:1996}. This assumption has proven to be adequate for many quiescent, 
old active regions and even for the non-eruptive global coronal-heliospheric interface \cite{Wang:1990,Schrijver:2003}. 

Studies of the coupling of the coronal field into the heliosphere suggest that the global coronal magnetic field is often 
largely potential. For the practical calculation of the global field, the so-called source-surface model has been introduced \cite{Schatten:1969}, 
in which the influence of the solar wind is artificially taken account of by the requirement that the field be radial at some 
exterior spherical (source) surface typically at $R_{s}=2.5R_{\odot}$ from the sun’s center. The potential-field source surface (PFSS) model, 
uses this concept to extrapolate the line-of-sight surface magnetic field through the corona with the boundary assumed to be at 
the source surface.

The potential field approximation has often been used to deduce the magnetic structure of the solar corona from measurements 
of the photospheric field. It obeys the equation $\nabla\times\textbf{B}=0 $, so that the magnetic field can be expressed as 
the gradient of a scalar potential, i.e. $\textbf{B}=-\nabla\Phi$. Since the magnetic field is divergence-free ($\nabla\cdot\textbf{B}=0 $), 
the scalar potential obeys the Laplace equation $\nabla^{2}\Phi=0$. Together with the photospheric boundary condition, which is 
traditionally provided by a map of the radial magnetic field, the Laplace equation can be solved in the spherical coordinate system 
$(r, \theta, \phi)$, where $\theta$ stands for colatitude. Using the radial magnetic field observed at $R_{\odot}$ and the source 
surface assumption at $R_{s}$ as the boundary condition, together with the spherical harmonic expansion in the domain $R_{\odot} < r < R_{s}$,
we finally obtain the potential field model.
%==============================================================================================================================================
\section{Results}
%==============================================================================================================================================
In this study, we have used potential field model and our spherical NLFFF optimization procedure to a group of active regions observed on 
November 09 2011 around 17:45UT by SDO/HMI instrument. During this observation there were four active regions (ARs 11338, 11339, 11341 and 11342, 
see Fig. \ref{fig2}a) along with other smaller sunspots spreading on the disk. In order to accommodate the connectivity between those ARs and their 
surroundings, and to study trans-equatorial loops (loops connecting north and south hemisphere), we adopt a non uniform spherical grid $r$, $\theta$, $\phi$ with 
$n_{r}=225$, $n_{\theta}=375$, $n_{\phi}=425$ grid points in the direction of radius, latitude, and longitude, respectively, with the field 
of view of 
$[r_{\mbox{min}}=1R_{\odot}:r_{\mbox{max}}=2R_{\odot}]\times[\theta_{\mbox{min}}=-50^{\circ}:\theta_{\mbox{max}}=50^{\circ}]\times[\phi_{\mbox{min}}=90^{\circ}:\phi_{\mbox{max}}=270^{\circ}]$. 
The main purpose of this work is to study the magnetic and electric current connectivities of the northern and southern solar hemispheres and to 
compare which of the two models (potential or NLFFF model) is best suited to fully describe those connectivities in full-disk environment.  

We use our preprocessing routine in spherical geometry to derive suitable boundary conditions for force-free modeling from the measured photospheric 
data \cite{Wiegelmann06sak,tilaye09}. We used a standard preprocessing parameter set $\mu_{1} = \mu_{2} = 1$, $\mu_{3}=0.001$ and $\mu_{4}=0.05$,  
which are similar to the values calculated from vector data used in previous studies \cite{Wiegelmann:2012,Tadesse:2013} for HMI data. The preprocessing 
influences the structure of the magnetic vector data. It does not only smooths $\textbf{B}_{t}$ (transverse field) but also alters its values in order 
to reduce the net force and torque. The change in $\textbf{B}_{t}$ is more pronounced than the radial component $\textbf{B}_{r}$ (radial field) since 
$\textbf{B}_{t}$ is measured with lower accuracy than the longitudinal magnetic field. 
%==============================================================================================================================================
\begin{figure}[htp!]
\begin{center}
     \includegraphics[viewport=5 0 558 795,clip,height=16.0cm,width=12.0cm]{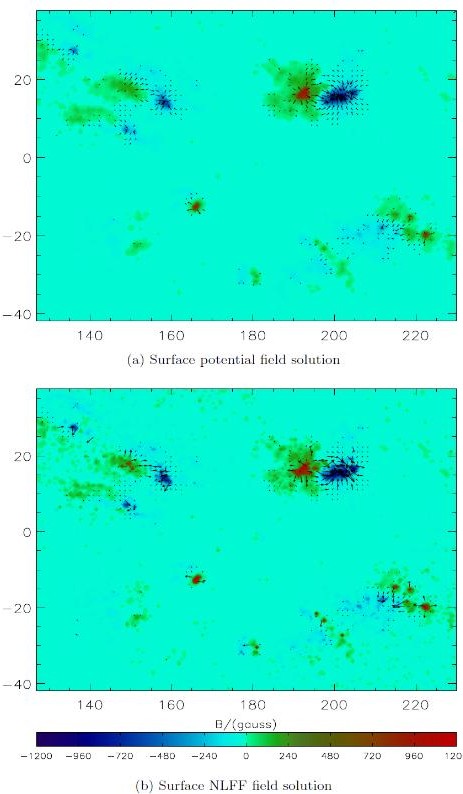}
     \end{center}
     \caption{Magnetic vector maps of HMI data on part of the lower boundary. The color coding shows $B_{r}$ on the photosphere 
and the black arrow indicates the transverse components of the field. The vertical and horizontal axes show latitude, $\theta$
and longitude, $\phi$ on the photosphere respectively.}
%\vspace*{1in}
\label{fig1}
\end{figure}
%==============================================================================================================================================

For computing the potential field, we use the preprocessed radial component $\textbf{B}_{r}$ of the HMI-data using spherical 
harmonic expansion. This potential field solution has been used to initialize our NLFFF code. For nonlinear force-free fields 
we minimize the functional Eq.~(\ref{4}). We implement the new term $L_{photo}$ 
in Eq.~(\ref{4}) to work with boundary data of different noise levels and qualities \cite{Wiegelmann10,Tilaye:2010}. For those pixels, 
for which $\textbf{B}_{obs}$ was successfully inverted, we allow deviations between the model field $\textbf{B}$ and the input fields 
observed $\textbf{B}_{obs}$ surface field using Eq.~(\ref{4}), so that the model field can be iterated closer to a force-free solution 
even if the observations are inconsistent. In order to control the speed with which the lower boundary is injected during the NLFFF 
extrapolation, we have used the Langrangian multiplier of $\nu=0.001$ as suggested by \inlinecite{Tadesse:2013}. For more details of 
the method used in this work we direct the readers to the study by \inlinecite{Tilaye:2010}. 

We plot the surface potential field and NLFFF solutions in Figure~\ref{fig1} for selected region excluding the polar regions. 
In this figure, one can see that there are substantial differences between the surface radial and transverse field components 
of the two models. To quantify the degree of disagreement between the respective vector components of potential and NLFFF model 
solutions on the bottom surface, we calculate their pixel-wise correlations. The correlation were calculated\cite{Schrijver06,Metcalf:2008} 
from
\begin{equation}
C_\mathrm{ vec}= \frac{ \sum_i \textbf{v}_{i} \cdot \textbf{u}_{i}}{ \Big( \sum_i |\textbf{v}_{i}|^2 \sum_i
|\textbf{u}_{i}|^2 \Big)^{1/2}}, \label{6}
\end{equation}
where $\textbf{v}_{i}$ and $\textbf{u}_{i}$ are the vectors at each grid point $i$ on the bottom surface. If the vector fields are 
identical, then $C_{vec}=1$; if $\textbf{v}_{i}\perp \textbf{u}_{i}$ , then $C_{vec}=0$. Table~\ref{table1} shows the correlation 
($C_{vec}$) of the 2D surface magnetic field vectors of potential and NLFFF models for the radial and transverse components. The vector 
correlation between $\textbf{B}_{t}$ of the potential and NLFFF surface vector maps is much more less than that of respective $\textbf{B}_{r}$ 
components. This is due to the fact that the transverse components ($\textbf{B}_{t}$) of the potential field model are computed using spherical 
harmonics expansion from the radial components of the magnetic field used as a boundary condition for NLFFF model. The average vector correlations 
between the radial and transverse components of the two models in Table~\ref{table1} indicates that the NLFFF solutions are somewhat far from 
potential as they carry electric currents.
\begin{table}
\caption{The correlations between the components of surface fields from potential and NLFFF models.}
\label{table1}
\begin{tabular}{ccc}
  \hline \hline
  v & u & $C_\mathrm{vec}$\\
\hline
$(\textbf{B}_{pot})_{r}$&$ (\textbf{B}_{NLFFF})_{r}$ &$0.937$\\
 $(\textbf{B}_{pot})_{t}$&$ (\textbf{B}_{NLFFF})_{t}$ &$0.813$\\
\hline
\end{tabular}
\end{table}
%==============================================================================================================================================
\begin{figure}[htp!]
   \centering
 \includegraphics[viewport=20 30 800 800,clip,height=12.2cm,width=12.0cm]{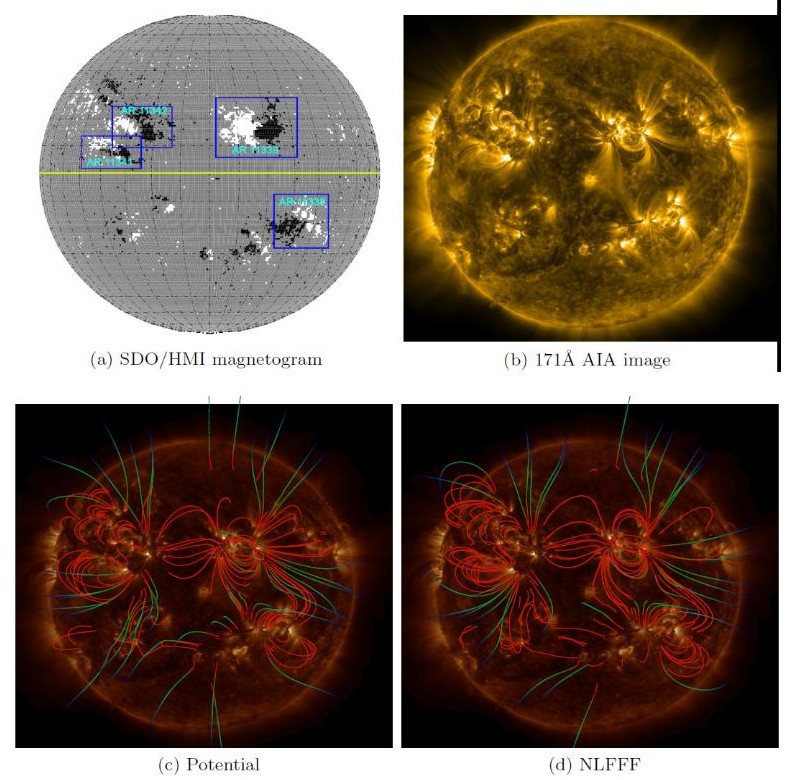}
\caption{a) Full-disk SDO/HMI magnetogram, the rectangles figure outline the main active regions during this observation 
and the yellow line indicates the solar equator ($\theta=0^{\circ}$), b) Full-disk AIA 171 \AA{} image. Both data sets were obtained 
on November 09, 2011 around 17:45UT, c) Field lines of the potential field model at 17:45 UT overlaid on AIA 193\AA{} image and d) Field lines of the 
NLFFF model at 17:45 UT overlaid on AIA 193\AA{} image. Green and red lines represent open and closed magnetic field lines.}\label{fig2}
 \end{figure}
%==============================================================================================================================================
In order to compare our reconstructions with observation, we plot the selected fieldlines of the potential and NLFFF models in 
Figure~\ref{fig2}c and d. We overlay the field lines with AIA 193 \AA{} image. For observation time in Figure~\ref{fig2}c and d, the field lines 
of the potential and NLFFF models are reconstructed from the same footpoints. The potential field lines in Figure~\ref{fig2}(c) have
an obvious deviation from the observed EUV loops, since the projection of the field lines at the bottom of AR 11339 leans toward the east, 
while the loops toward the west. The loops connecting positive and negative polarities of AR 11341 are best overlaid by NLFFF lines than 
potential ones. The spatial correspondence between the overall shape of the NLFFF field lines and the EUV loops is much
improved as shown in Figure~\ref{fig2}(d). Therefore, the qualitative comparison between the model magnetic field lines and the
observed EUV loops indicates that the NLFFF model provides a more consistent field for full-disk magnetic field reconstruction. 
In the absence of a more reliable quantitative comparison, it remains the best option even if the qualitative nature is not ideal.

In addition to the above comparison to quantify the degree of disagreement between vector field solutions of the two models in the 
computational volume that are specified on identical sets of grid points, we use five metrics that compare either local characteristics 
(e.g., vector magnitudes and directions at each point) or the global energy content in addition to the force and divergence integrals as 
defined in  \inlinecite{Schrijver06}. The vector correlation ($C_{vec}$) metric of Equation~(\ref{6}) is also used analogous 
to the standard correlation coefficient for scalar functions. The second metric, $C_{CS}$ is based on the Cauchy-Schwarz 
inequality($|\textbf{a}\cdot \textbf{b}\leq|\textbf{a}||\textbf{b}|$ for any vector $\textbf{a}$ and $\textbf{b}$)
\begin{equation}
C_{\rm CS} = \frac{1}{N} \sum_i \frac{{\bf B_i} \cdot {\bf b_i}} {|{\bf B_i}||{\bf
b_i}|},\label{7}
\end{equation}
where $\textbf{B}=\textbf{B}_{\mbox{pot}}$, $\textbf{b}=\textbf{B}_{\mbox{NLFFF}}$, and $N$ is the number of vectors in the field. 
This metric is mostly a measure of the angular differences of the vector fields: 
$C_{CS} = 1$ when $\textbf{B}$ and $\textbf{b}$ are parallel and $C_{CS} = -1$ if they are anti-parallel; $C_{CS} = 0$  if  
$\textbf{B}_{i}\perp \textbf{b}_{i}$ at each point. Next, we introduce two measures for the vector errors, one normalized to 
the average vector norm, one averaging over relative differences. The normalized vector error $E_{N}$ is defined as
\begin{equation}
E_{\rm N} = \sum_i |{\bf b_i}-{\bf B_i}|/ \sum_i |{\bf B_i}|,\label{8}
\end{equation}
The mean vector error $E_{M}$ is defined as
\begin{equation}
E_{\rm M} = \frac{1}{N} \sum_i \frac{|{\bf b_i}-{\bf B_i}|}{|{\bf B_i}|}.\label{9}
\end{equation}
Unlike the first two metrics, perfect disagreement of the two vector fields results in $1-E_{M }= 1-E_{N} = 0$. As we are also 
interested in determining how well the models estimate the energy contained in the field, we use the total magnetic energy in 
NLFFF model field normalized to the total magnetic energy in potential field as a global measure of the quality of the fit:
\begin{equation}
\epsilon = \frac{\sum_i |{\bf b_i}|^2}{\sum_i |{\bf B_i}|^2}.\label{10}
\end{equation}
$\epsilon=1$, if there is no difference between the potential field and the nonlinear force-free model solutions. 
 %==============================================================================================================================================
\begin{table}
\caption{Comparision of figure of merits between potential field and NLFFF models. We have used pherical grids of $225 \times 375 \times 425$.}
\label{table2}
\centering
\begin{tabular}{ccccccccc}   
\hline \hline Model &$L_{f}$&$L_{d}$ &$C_{\rm vec}$&$C_{\rm CS}$&$1-E_{N}$&$1-E_{M}$&$\epsilon$& Time \\
\hline
%&\multicolumn{3}{c}{Spherical grid $225 \times 375 \times 425$} &&&& \\
Potential &$0.000$& $ 0.002$&$ 1$&$ 1$&$ 0$&$ 0$&$1 $&30min  \\
 NLFFF &$0.591$&$0.997$&$0.878$&$0.833$&$0.735$&$0.876$&$1.220$&4h:39min \\
\hline
\end{tabular}
\end{table}
%==============================================================================================================================================
The degree of convergence towards a force-free and divergence-free model solution can be quantified by the integral 
measures of the Lorentz force and divergence terms in the minimization functional in Equation (\ref{4}), computed 
over the entire model volume V. $L_{f}$ and $L_{d}$ of Equation (\ref{4}) measure how well the force-free and divergence-free 
conditions are fulfilled, respectively. In Table~\ref{table2}, we provide some quantitative measures to rate the quality 
(figure of merit) of our reconstruction. Column $1$ names the corresponding models. Columns $2-3$ show how well the force 
and solenoidal condition are fulfilled for both models. The next five columns of Table~\ref{table2} contain different measures 
which compare our NLFFF solution with potential field. Those figures of merit indicate that there is a clear difference between 
potential field and NLFFF models solutions for full-disk. The last column shows the computing time on $1$ processor. The figures 
of merit show that the potential field is far away from the true solutions and contains only $81.97\%$ of the total magnetic energy.

\begin{figure}[htp!]
   \centering

     \includegraphics[bb=10 0 740 800,clip,height=11.5cm,width=11.5cm]{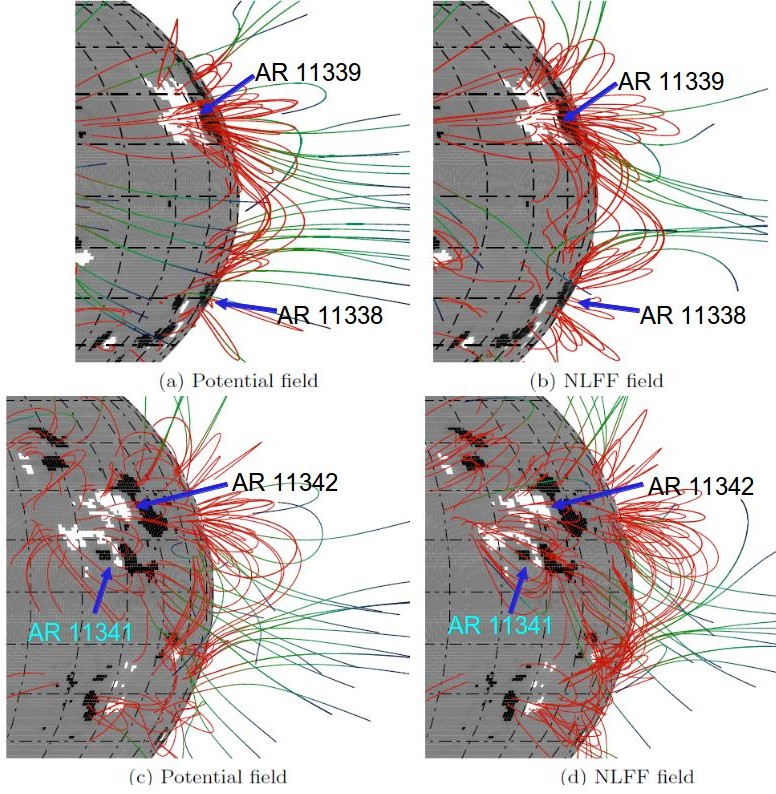}
\caption{a) Selected potential field lines connecting ARs 11338 and 11339 rotated to limb.  b) The corresponding field lines for NLFFF in a), 
c) Selected potential field lines connecting AR 11341 and Sun spots in southern hemisphere rotated to limb d) The corresponding 
field lines for NLFFF in c).}\label{fig3}
 \end{figure}

\begin{figure}[htp!]
 \includegraphics[viewport=5 0 558 795,clip,height=16.0cm,width=12.0cm]{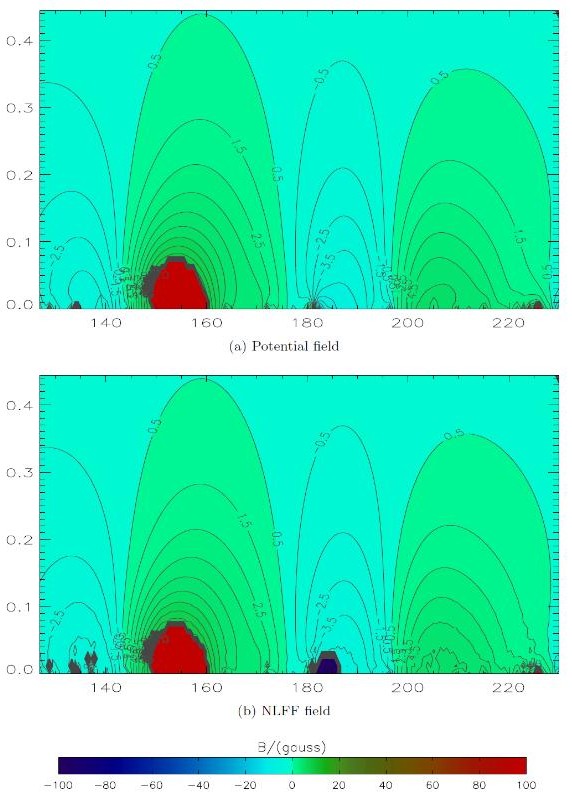}
     \caption{Surface contour plot of perpendicular component of magnetic fields crossing a plane of the solar equator ($\theta=0^{\circ}$). 
The color coding shows $B_{\perp}$ crossing the $\theta=0^{\circ}$ equatorial plane. The vertical and horizontal axes show height, $r$(in solar radius) 
and longitude, $\phi$(in degree) on the plane.}
\label{fig4}
\end{figure}

\begin{figure}[htp!]
\includegraphics[viewport=5 0 558 795,clip,height=16.0cm,width=12.0cm]{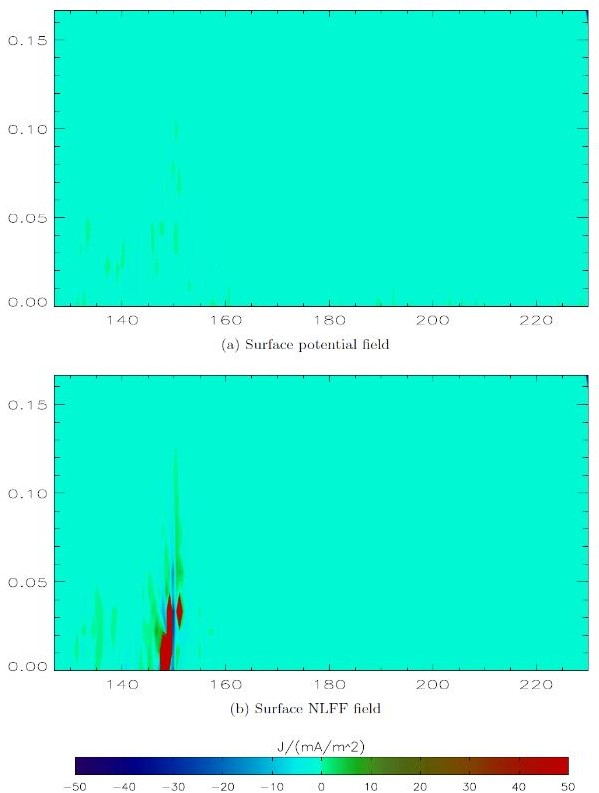}
     \caption{Surface contour plot of perpendicular component of electric current density crossing a plane of the solar equator 
($\theta=0^{\circ}$). The color coding shows $J_{\perp}$ crossing the $\theta=0^{\circ}$ equatorial plane. The vertical and horizontal 
axes show height, $r$(in solar radius) and longitude, $\phi$(in degree) on the plane.}
\label{fig5}
\end{figure}

We estimate the free magnetic energy to be the difference between the extrapolated NLFFF and the potential field 
with the same normal boundary conditions in the photosphere\cite{Regnier,Thalmann}. We therefore estimate the upper limit to 
the free magnetic energy associated with coronal currents of the form
\begin{equation}
E_\mathrm{free}=\frac{1}{8\pi}\int_{V}\Big(B_{nlff}^{2}-B_{pot}^{2}\Big)r^{2}sin\theta dr d\theta d\phi, \label{ten}
\end{equation}
Our result for the estimation of free-magnetic energy in Table~\ref{table3} shows that the NLFFF model has $18.83\%$ more energy 
than the corresponding potential field model. Therefore, whenever we try to analysis magnetic field configuration and its magnetic 
field contents in full-disk, it is inaccurate to use potential field model as it undermines the real features of the field in the corona.
%==============================================================================================================================================
\begin{center}
\begin{table}
\caption{The magnetic energy  associated with extrapolated potential and NLFFF field configurations from full disk SDO/HMI data.}
\label{table3}
\begin{tabular}{ccc}
 \hline \hline
Model & $E_{\mbox{total}}(10^{33}{\mbox{erg}})$& $E_{\mbox{free}}(10^{33}{\mbox{erg}})$\\
\hline
Potential &$7.306$&$0$\\
NLFFF &$8.913$&$1.607$\\
\hline
\end{tabular}
\end{table}
\end{center}
%==============================================================================================================================================
In our previous work \cite{Tilaye:2012a}, we have studied the connectivity between three neighbouring active regions. 
In this work, we study the magnetic and electric connectivities between active regions in northern and southern solar 
hemispheres using full-disk HMI data. During this observation, there were three active regions in the northern hemisphere 
and one active region with surrounding sunspots in the southern hemisphere. As a result the total unsigned magnetic field 
flux in the northern hemisphere is larger than that of the southern one.  In order to quantify these connectivities, we 
have calculated the magnetic flux and the electric currents shared between those active regions. For the magnetic flux, 
e.g., we use
\begin{equation}
\Phi_{\alpha\beta}=\sum_{i}|\textbf{B}_{i}\cdot\hat{r}|R^{2}_{\odot}\textrm{sin}(\theta_{i})\Delta\theta_{i}\Delta\phi_{i}\label{ten}
\end{equation}
where the summation is over all pixels of $\mbox{AR}_{\alpha}$ from which the field line
ends in $\mbox{AR}_{\beta}$ or $i\in\mbox{AR}_{\alpha}\|\,\mbox{conjugate footpoint}(i)\in\mbox{AR}_{\beta}$.
The indices $\alpha$ and $\beta$ denote the active region. For the electric current we 
replace the magnetic field, $\textbf{B}_{i}\cdot\hat{r}$, by the vertical current density $\textbf{J}_{i}\cdot\hat{r}$ 
in Equation (\ref{ten}). Table~\ref{table4} shows the percentage of the total magnetic flux and electric current shared 
between the ARs 11339 and 11338 and between AR 11341 and sunspots in the southern hemisphere (see Fig.~\ref{fig3}). We have 
calculated total an unsigned flux for each active regions 11339 and 11341 (both being in northern hemisphere) and the flux due to 
those field lines ending in the AR 11338 and magnetic patches in the southern hemisphere, respectively. The first column of 
Table~\ref{table4} shows that $7.31\% (6.17\%)$ of positive/negative polarity of AR 11339 (11341) in the northern hemisphere 
is connected to positive/negative polarity of AR 11338 (magnetic patches) in southern hemisphere for potential field configuration. 
The percentage share between those ARs are $7.57\% (6.91\%)$ for NLFFF configuration. In Table~\ref{table4}, the percentages in 
the bracket show the share in electric current between the corresponding active regions and small magnetic patches. 
Figure~\ref{fig3} display the trans-equatorial loops connecting those active regions. In the figure, we rotate those active 
regions to the limb to show loops connecting the two hemispheres. The surface contour plots in Figures~\ref{fig4} and \ref{fig5} show 
the magnetic flux and electric current density flux of field lines crossing equatorial plane, respectively. From those figures we can see 
that NLFFF model has more magnetic and electric current density fluxes compared to the potential field. Figure~\ref{fig5} shows that there 
are more electric currents crossing the plane carried by trans-equatorial loops connecting AR 11341 in northern hemisphere to sunspots 
in the south than those currents connecting ARs 11339 and 11338. This indicate that much of the trans-equatorial loops connecting ARs 
11339 and 11338 are potential. In general the two hemispheres are more magnetically connected than electric current. However, one has 
to use NLFFF model over potential model to study magnetic field structure in any phenomena involving the field.   
%==============================================================================================================================================
%==============================================================================================================================================
%==============================================================================================================================================
%%==================================================================================================================
\begin{center}
\begin{table}
\caption{The percentage of the total magnetic flux and electric current density shared between the ARs 11339 and 11338 and 
between AR 11341 and sunspots in the southern hemisphere ( see Fig.~\ref{fig3}). The percentage in brackets denote that of 
electric currents.}
\label{table4}
\begin{tabular}{ccc}
 \hline \hline
Model & $\mbox{AR 11339} \longmapsto \mbox{AR 11338}$& $\mbox{AR 11341} \longmapsto \mbox{Patches in South}$\\
\hline
Potential &$7.31\,(0.0)$&$6.17\,(0.01)$\\
NLFFF &$7.57\,(0.13)$&$6.91\,(3.35)$\\
\hline
\end{tabular}
\end{table}
\end{center}
%==============================================================================================================================================
\section{Conclusion and outlook}
%==============================================================================================================================================
Both potential and nonlinear force-free field (NLFFF) codes in Cartesian geometry are not well suited for larger domains, since the spherical 
nature of the solar surface cannot be neglected when the field of view is large. One of the most significant results of Solar Dynamic Obseratory 
(SDO) mission to date has been repeated observations of large, almost global scale events in which large scale connection between active 
regions may play fundamental role. Therefore, it appears prudent to implement a NLFFF procedure in spherical geometry for use when large scale 
boundary data are available, such as from the Helioseismic and Magnetic Imager (HMI) on board SDO. 

In this study, we have investigated the coronal magnetic field associated with full solar disk on November 09 2011 by analysing SDO/HMI data using 
potential and NLFFF models. During this particular observation, there were three active regions in the northern hemisphere and one active region 
surrounded by sunspots in the south. We have used our spherical NLFFF and potential codes to compute the magnetic field solutions in full-disk. 
For computing the potential field, we use the preprocessed radial component $\textbf{B}_{r}$ of the HMI-data using spherical harmonic expansion. 
We implement our NLFFF code initialized by the this potential field solution (except the lower observed bottom boundary) during relaxation towards 
force-freeness state in the computational volume.

We have compared the magnetic field solutions from both models. The qualitative comparison between the model magnetic field lines and the 
observed EUV loops indicates that the NLFFF model provides a more consistent field for full-disk magnetic field reconstruction. In addition 
to this, the figures of merits have been used to quantify the disagreements between the two models field lines. The figures of merit indicate 
that the magnetic field lines structures are not potential in full-disk field-of-views. However, much of the trans-equatorial field lines 
connecting the active regions in the northern and southern hemispheres are potential. This indicates that the two solar hemispheres are 
more magnetically connected than electric current. The magnetic field lines obtained from nonlinear force-free extrapolation bear larger 
total magnetic energy that of the potential field model. Therefore, one has to use NLFFF model over potential model to study magnetic field 
structure in any phenomena involving the field in full-disk.    
%==============================================================================================================================================

\section*{Acknowledgements} %The authors thank the anonymous referee for helpful comments. 
Data are courtesy of NASA/SDO and the AIA and HMI science teams. SOLIS/VSM vector magnetograms are produced cooperatively by NSF/NSO and NASA/LWS. 
The National Solar Observatory (NSO) is operated by the Association of Universities for Research in Astronomy, Inc., under cooperative agreement 
with the National Science Foundation. This work was supported by NASA grant NNX07AU64G and the work of T. Wiegelmann was supported by DLR-grant $50$ OC $453$  $0501$.
%==============================================================================================================================================

\bibliographystyle{spr-mp-sola}
\bibliography{bibtex}

\end{article}
\end{document}